\documentclass[11pt]{article}

\usepackage[final]{acl}

\usepackage{times}
\usepackage{latexsym}
\usepackage{subcaption}
\usepackage{listings}
\usepackage{graphicx}
\usepackage{booktabs}
\usepackage{amssymb}
\usepackage{pifont}
\usepackage{multirow}
\usepackage{colortbl}
\usepackage[nodisplayskipstretch]{setspace} 
\newcommand{\cmark}{\ding{51}}%
\newcommand{\xmark}{\ding{55}}
\usepackage{colortbl}
\usepackage[T1]{fontenc}

\usepackage[utf8]{inputenc}

\usepackage{microtype}

\usepackage{inconsolata}

\newcommand{\ctslogo}{\raisebox{3.4pt}{\includegraphics[scale=0.01]{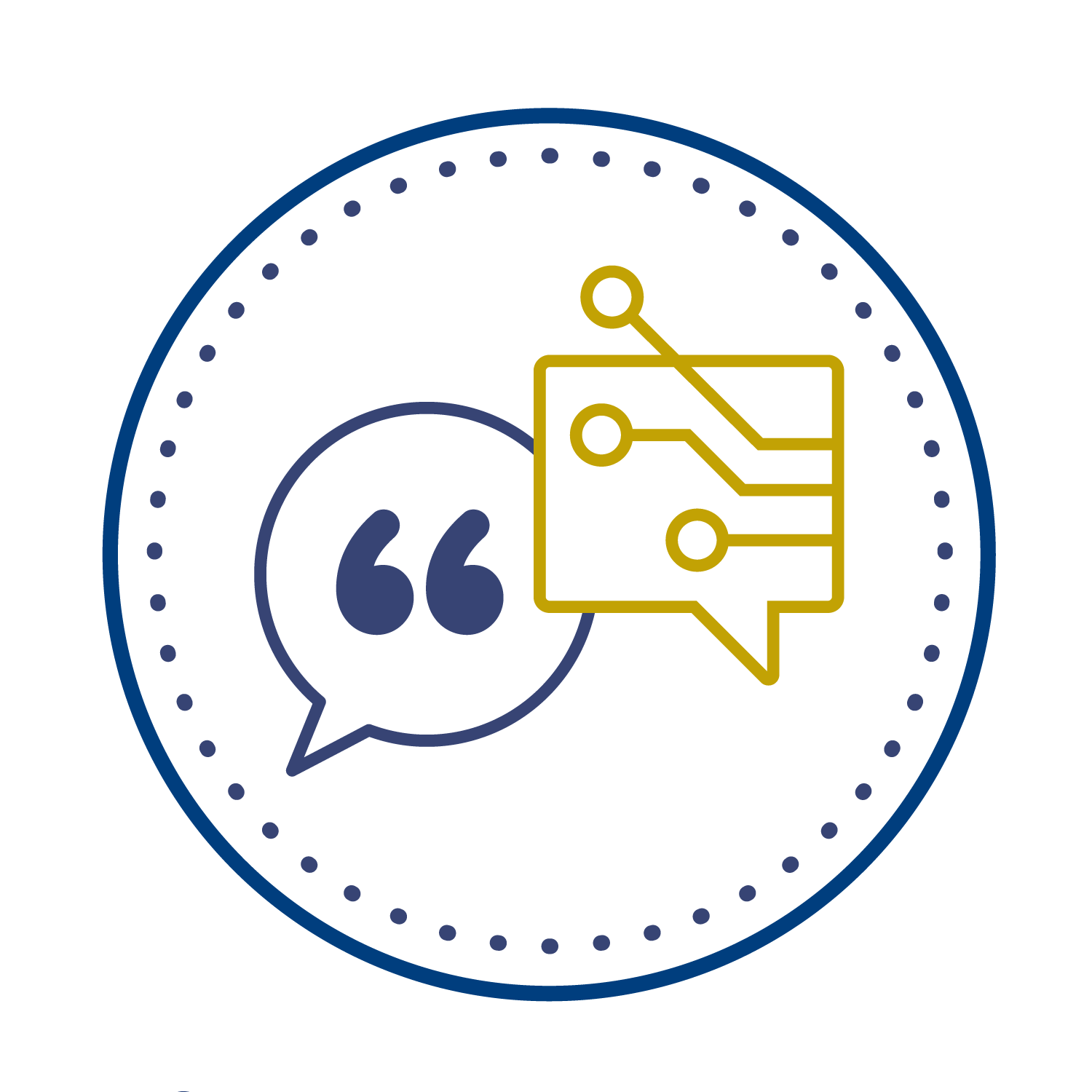}}}
\newcommand{\PAIlogo}{\raisebox{3.4pt}{\includegraphics[scale=0.08]{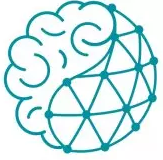}}}
\newcommand{\ebaylogo}{\raisebox{3.4pt}{\includegraphics[scale=0.02]{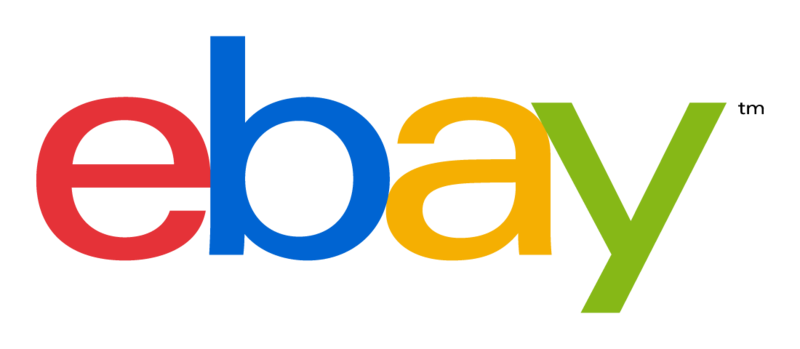}}}

\title{Centrality-aware Product Retrieval and Ranking}

\author{
    Hadeel Saadany\ctslogo, Swapnil Bhosale\PAIlogo, Samarth Agrawal\ebaylogo, \\
    \bf{Diptesh Kanojia\PAIlogo, Constantin Or\u{a}san\ctslogo, and Zhe Wu\ebaylogo}\\ [.35em]
    \PAIlogo Institute for People-Centred AI, University of Surrey, United Kingdom\\
    \ctslogo Centre for Translation Studies, University of Surrey, United Kingdom\\ [.15em]
     \ebaylogo eBay Inc., USA\\ [.4em]
    \texttt{\{hadeel.saadany, s.bhosale, d.kanojia, c.orasan\}@surrey.ac.uk},\\ 
    \texttt{\{samagrawal,zwu1\}@ebay.com}
}

\begin{document}
\maketitle
\begin{abstract}
This paper addresses the challenge of improving user experience on e-commerce platforms by enhancing product ranking relevant to users' search queries. Ambiguity and complexity of user queries often lead to a mismatch between the user's intent and retrieved product titles or documents. Recent approaches have proposed the use of Transformer-based models, which need millions of annotated query-title pairs during the pre-training stage, and this data often does not take user intent into account. To tackle this, we curate samples from existing datasets at eBay, manually annotated with \textit{buyer-centric relevance} scores and \textit{centrality} scores, which reflect how well the product title matches the user’s intent. We introduce a User-intent Centrality Optimization (UCO) approach for existing models, which optimises for the user intent in semantic product search. To that end, we propose a dual-loss based optimisation to handle hard negatives, \textit{i.e.,} product titles that are semantically relevant but do not reflect the user's intent. Our contributions include curating challenging evaluation sets and implementing UCO, resulting in significant product ranking efficiency improvements observed for different evaluation metrics. Our work aims to ensure that the most buyer-centric titles for a query are ranked higher, thereby, enhancing the user experience on e-commerce platforms.
\end{abstract}

\section{Introduction}
Achieving a user-focused experience on e-commerce platforms (eBay, Walmart, Amazon, Etsy, JD) is enabled by ranking products relevant to the user's intent expressed via the search query. However, user queries often do not fully reflect the underlying intent behind the search terms used within the query. For example, ambiguous queries like \textit{`iphone 13'}, or \textit{`i5 pc 1tb 16gb 8gb gpu'} can lead to many variants. To aggravate the challenge further, user queries can consist of lexical terms with alphanumeric characters, which do not reveal a semantic match within existing product titles. Information Retrieval (IR) systems depend upon semantic similarity/distance between words or phrases used in the search query and the product title. Therefore, ranking the product titles based on only lexical or only semantic query-title match can be a particularly challenging problem, as detailed in the examples below: 

\paragraph{Ambiguous Queries} Some queries can be ambiguous and do not clearly reflect the user's intention. From the same example above, for a query like \textit{`iPhone 13'}, the user is most likely looking to buy the base variant or to check out other device variants. However, this intent is not clear from the query, and the system can even rank `iPhone 13 cover' among the top retrieved products. Hence, a major challenge faced by search systems is to retrieve titles that are likely to be relevant to the user intent at high ranks, and push down negative titles such as `iPhone 13 cover' which have semantic proximity to positive titles within the embedding space of the computational model but may not reflect user’s underlying objective.
\paragraph{Repetition} Similar to the example above, the repetition of the exact string of words from a user's query, such as `iPhone 13', in both relevant and irrelevant titles often renders embeddings-based similarity approaches futile as the proximity of positive and negative titles in the embedding space may not be reflective of their relevancy. In such cases, human annotation towards user intent for a query-title pair is needed to establish a clear ranking among products retrieved by the model.
\paragraph{Alphanumeric Queries} Queries such as \textit{`S2716DG'} consist of alphanumeric characters where a letter or number can signify important detail for the product/model. For example, based on the naming convention of PC monitors, a single letter defines the type of panel in the product. In this case, the Dell S2716DG is a 27-inch monitor with a TN panel, and changing the last letter to P would refer to a monitor with an IPS panel. Similarly, product colour or a specific spare part can be identified from such queries. Unless the product title contains this alphanumeric sequence of characters, the semantic similarity between the query and a non-intended product can be high, thus misleading the system.

\begin{figure}[!t]
   \begin{minipage}{0.48\textwidth}
     \centering
     \includegraphics[width=.55\linewidth]{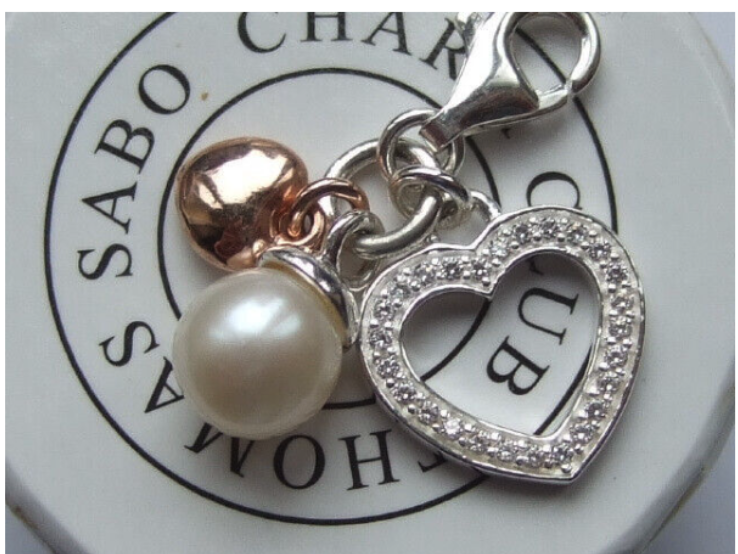}
     \caption{Central Title: Thomas sabo charms with \\ 18k Rose gold pearl}
     \label{fig:central}
   \end{minipage}\hfill
   \begin{minipage}{0.48\textwidth}
     \centering
     \includegraphics[width=.55\linewidth]{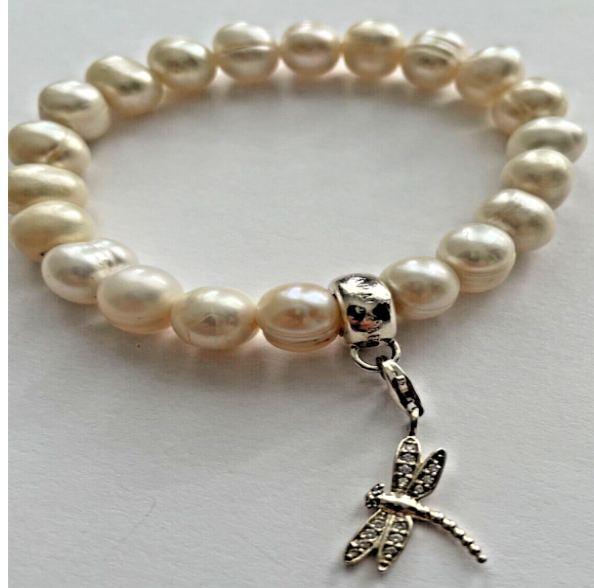}
     \caption{Non-central title: Thomas Sabo charm \\club bracelet with detachable dragonfly charm}
     \label{fig:non-central}
   \end{minipage}
\end{figure}

In this paper, we investigate the challenges listed above and take a two-step approach to improve product retrieval and ranking. We curate samples from existing internal datasets at eBay consisting of user search queries paired with retrieved product titles on their platform. These datasets are human-annotated based on detailed guidelines to produce two buyer-centric relevance annotations. First, a widely used relevance ranking schema where query-title pairs are provided a ranked class from among Bad (1), Fair (2), Good (3), Excellent (4) and Perfect (5), where `\textit{perfect}' reflects an exact query-title pair match, \textit{i.e.,} the annotator is very confident that the user found precisely what they were looking for, while `\textit{bad}' reflects no match between the product and the need expressed in the query \cite{10.1145/3326937.3341259,KANG201653}. Second, query-title pairs are annotated with a binary \textit{centrality score}, obtained from majority voting over multiple human annotations, \textit{i.e.,} indicating whether the item reflects the need expressed in the query.  The difference between centrality and relevance scoring is that the latter detects whether an item is an outlier, a surprising addition to the recall set, or the item centrally matches the expectations.
Figures \ref{fig:central} and \ref{fig:non-central} show two examples of the centrality annotation for the same query, ``Thomas Sabo charm''. Figure \ref{fig:central} shows a product central to the query since, based on purchase data, this query typically reflects the user's need for a charm (a small ornament worn on a necklace or bracelet). On the other hand, the product in Figure \ref{fig:non-central} is not central to the user's intent as it is a \textit{Thomas Sabo charm} attached to a bracelet; the user intent is a charm, not a bracelet. Although both titles are semantically related to the query, based on the \textit{degree of specificity} expressed in the query, the product in Figure \ref{fig:non-central} becomes less central to the user's intent and gets annotated with 0 as its centrality score whereas the product in Figure \ref{fig:central} receives 1. We use an internal human-annotated dataset for this task. Henceforth, we refer to it as Internal Graded Relevance or IGR dataset.

We extract challenging evaluation sets from the IGR dataset based on the challenges discussed above. Our objective is to increase the retrieval and ranking efficiency of product search by training a model for query-title pairs that integrates the user intent in the similarity algorithm. Given the search query, we propose using a user-intent centrality optimisation (UCO) step for existing models which cater to the ranking of relevant products. Further, we propose utilising a dual-loss based optimisation to address the query-title pairs which constitute hard negatives, \textit{i.e.,} query-title pairs where the product title is semantically relevant to the user's query but is annotated as non-central to the user intent, or has \textit{Bad} or only \textit{Fair} annotated relevancy.


We hypothesise that there is an unwanted semantic proximity of such negative titles to their search queries in the model embeddings space. To improve search, we optimise the existing ranking model with our dual-loss-based optimisation approach, ensuring that the retrieval algorithm should have the most ``typical'' titles for a query ranked highly than other titles which \textit{may be relevant but are not typical}. Our contributions are 1) curating challenging evaluation sets that cater to this problem and 2) user-intent centrality optimisation (UCO), which results in a stark improvement on all the evaluation sets. 
\section{Related Work}

Our work is based on a two-step approach to improve product ranking given a search query for retrieving items. Existing literature on traditional candidate retrieval research focused on learning query rewrites~\cite{bai2018scalable,guo2008unified} as an indirect approach to bridge the vocabulary gap between queries and documents/titles. Some approaches, including latent semantic indexing with matrix factorization~\cite{deerwester1990indexing}, and with probabilistic models~\cite{hoffman1990probabilistic}, and semantic hashing with an auto-encoder~\cite{salakhutdinov2009semantic}, have been proposed. Most of these are unsupervised models based on word co-occurrence in documents/product titles.

Modern IR systems deploy semantic retrieval models as bi-encoders~\cite{muennighoff2022sgpt} or Siamese networks~\cite{chiang-chen-2021-bert} comprising two encoders. Most existing studies focus on designing or pre-training encoders with different representation learning approaches~\cite{gao2011clickthrough,salakhutdinov2009semantic,yih2011learning,huang2020embedding,liu2020decoupled}. Representative works, namely, the Deep Semantic Similarity Model (DSSM)~\cite{huang2013learning}, and CDSSM~\cite{shen2014learning}, are some of the earliest methods which utilise a deep neural network (DNN) using clickthrough data. Subsequently, CNNs~\cite{gao2014modeling,shen2014latent,shen2014learning,severyn2015learning} and RNNs~\cite{palangi2014semantic,palangi2016deep} have been utilised for semantic retrieval. Recently, new models, including DRRM~\cite{guo2016deep} and Duet~\cite{mitra2017learning} were developed to include traditional IR lexical matching (\textit{e.g.,} exact matching, term importance) within semantic retrieval performed by DNNs. However,~\cite{mitra2018introduction} argues that most works proposed in this direction focus on the ranking stage, where the optimisation objectives differ from candidate title retrieval. To further improve the performance of semantic retrieval, Transformer-based Pre-trained Language models (PTLMs) like BERT~\cite{devlin2018bert} and ERNIE~\cite{zhang2019ernie} have been leveraged~\cite{fuchs2020intent,wang2024utilizing,liu2021pre}. Using larger pre-trained models, semantic retrieval has observed a significant performance improvement and generalisation for retrieval but without a specific focus on ambiguous or alphanumeric queries, which is what we essentially address in this paper.

Further, interaction-based approaches~\cite{moe2003buying,long2012enhancing,gu2020deep,yates2021pretrained,zou2020neural,dai2023contrastive} have also been widely used for IR systems, which further go into semantic matching to model for query-document/title interaction using DNNs~\cite{lu2013deep,mitra2017learning,wan2016deep,zhao2020memory,kabir2022ordsim}. Most of these approaches focus on user personalisation needs, and often rely on hand-crafted rules. Often, such approaches cannot cache the document embeddings offline for faster retrieval, and may be inefficient for retrieval~\cite{liu2021pre}. ~\cite{su2018user} use the results of an online survey and search logs from a commercial product search engine to show that product search falls into categories like Target Finding, Decision Making and Exploration.~\cite{10.1145/3470564} propose Personal Word-embeddings for Personalized Search (PEPS) which uses as additional layer trained on user embeddings and personal logs. 

While personalised embeddings and interaction-based approaches improve ranking performance for ambiguous user queries, our work focuses on dealing with similar challenges using a different approach \textit{infusing centrality-awareness}. To be considered an impactful solution for the challenges at hand, we believe that product ranking approaches can be more generalised compared to personalised embeddings, improving the base retrieval with a focus on user intent. Our approach utilises two existing loss functions that cater to the task and optimise the retrieval model, which can be used at both stages, retrieval and ranking.

\section{Methodology}

\subsection{Baseline Model: eBERT}

For training our system, we employ the in-house multilingual eBERT\footnote{\href{https://innovation.ebayinc.com/tech/engineering/how-ebay-created-a-language-model-with-three-billion-item-titles/}{eBERT Language Model}} model. eBERT is trained on item/product data from eBay and general domain (Wikipedia and RefinedWeb) text. The item data used to train this model consists of approximately $3$ billion item titles. We also test another eBERT variant, eBERT-siam, which is fine-tuned to generate similar embeddings for item titles using a Siamese network. This model is designed specifically for tasks related to similarity search on query and product titles. Both models are used offline to perform experiments and are optimised with UCO to note performance changes for retrieval and ranking.
\subsection{\mbox{User-intent Centrality Optimization (UCO)}}

We perform UCO as an optimisation step to overcome the problem of top-ranked, hard negative query-title pairs that are semantically relevant but not central to user intent. Thus, we fine-tune the baseline model with a supervised binary classification task on product centrality. Then, based on our hypothesis for transfer learning capabilities, we employ the knowledge learned from the domain information of centrality optimisation as an inductive bias to boost the ranking capability of a retrieval model, thereby, optimising the ranking task for our challenging evaluation sets. We employ dual-loss optimisation, as explained in the next section.

\begin{figure}[t!]
    \centering
        \includegraphics[width=\columnwidth,trim={0.1cm .2cm .1cm 0cm},clip]{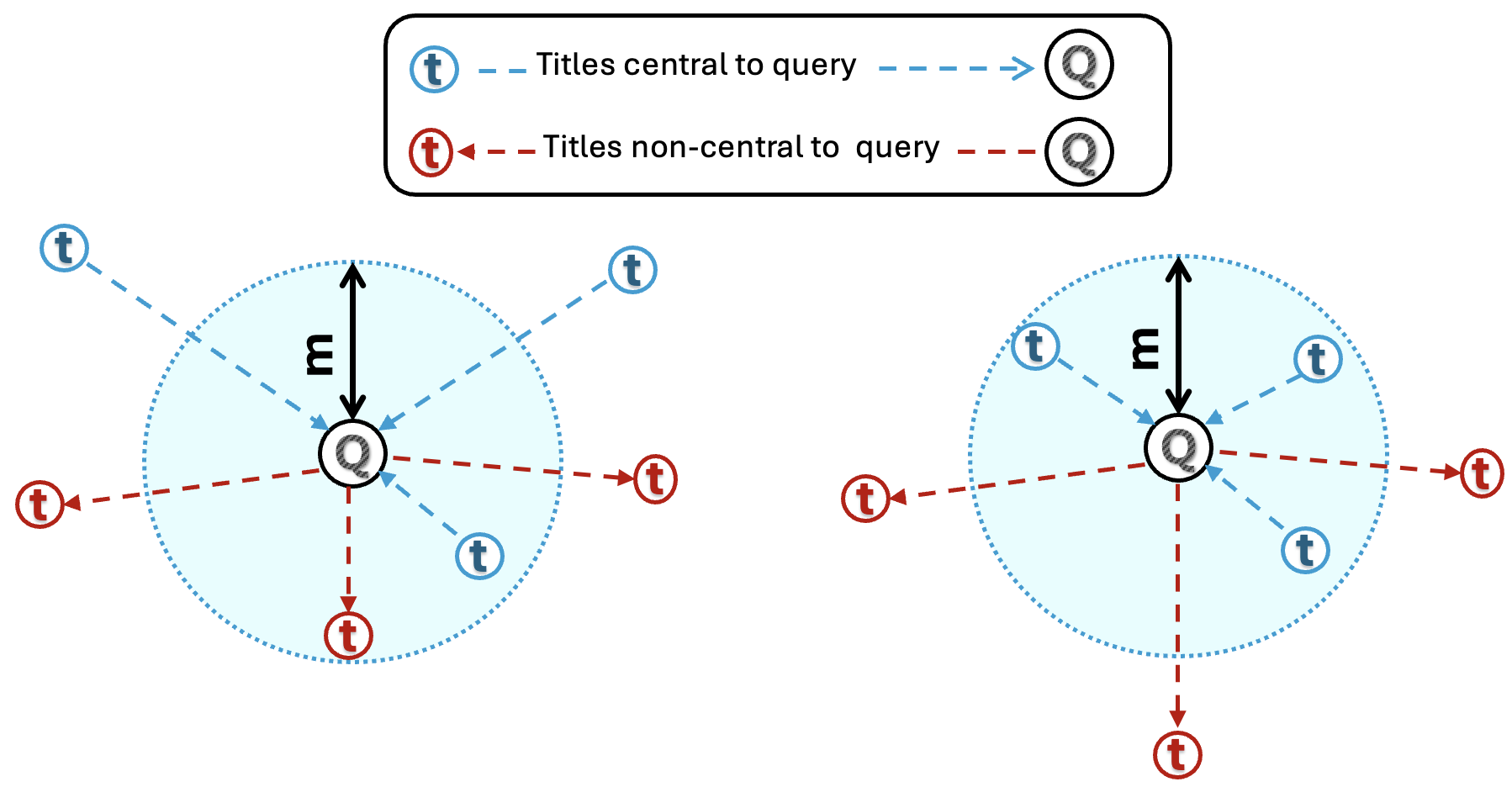}
        \caption[Caption for LOF]{The figure shows how the loss function algorithm works with hard negatives. The algorithm targets those non-central titles (red) that are inside the margin.\footnotemark}
        \label{fig:loss}
\end{figure}
\footnotetext{Adopted from \cite{hadsell2006dimensionality} with modifications.}

\subsection{Dual-Loss Based Optimisation}
 \textbf{Multiple Negative Ranking Loss} (MNRL) \cite{henderson2017efficient} is the first loss function we employ. MNRL quantifies the difference between positive and negative samples for a query. MNRL is used to create a clear distinction between relevant (positive) and irrelevant (negative) data points, achieved by minimising the distance between the query and positive samples while maximising it for multiple negative samples. Multiple negatives provide more context, enabling the optimisation to discriminate between varying degrees of irrelevance. Mathematically, it can be represented as follows:
\begin{equation}
\resizebox{\columnwidth}{!}{
MNRL = $\sum_{i=1}^P \sum_{j=1}^N max(0,f(q,p_i) - f(q,n_j) + margin)$
}
\end{equation}

\noindent where $P$ is the number of positive titles, $N$  is the number of negative titles, $q$ is the query, $f$ is our similarity function, which is cosine similarity, and $margin$ is a hyperparameter defining the optimum distance between positive and negative titles defined by the centrality of the user-intent. The MNRL minimises the distance between $(q,p_i)$ while it simultaneously maximises the distance $(q,n_j)$ for all $P$ and $N$ titles.

\textbf{Online Contrastive Loss (OCL)} is a variant of Contrastive Loss (CL) \cite{carlsson2020semantic}. 
OCL attends to negative pairs that have a lower distance than the positive pairs with the largest distance, as well as, the positive pairs that have a higher distance than the lowest distance of negative pairs, \textit{i.e.,} the hard cases in a batch, and computes the loss only for these cases. It selects hard positive (positives that are far apart) and hard negative pairs (negatives that are close), and backpropagates only for such pairs. OCL can be represented as follows:
\begin{equation}
\resizebox{0.83\columnwidth}{!}{OCL = $Y * D + (1-Y) * max(margin-D, 0)^2$}
\end{equation}

\noindent where $Y$ is our centrality score between the query and title, it will be $1$ if the title is central to the user intent and $0$ if it is not. The $D$ variable is the function that returns the distance between the query and title embeddings, which is the cosine similarity in our case. The $max$ function takes the largest value of $0$ and the $margin$ minus the distance. The negative samples (centrality = $0$) should have a distance of at least the margin value which we empirically set during training. This means that if we define some radius/margin, all the central titles should fall inside this margin, and all the non-central ones should fall outside.

MNRL primarily reduces the distance between positive pairs out of a large set of possible candidates and hence works particularly well when the dataset has a significant number of positives, which caters to the dataset skew in our case. However, MNRL does not push dissimilar pairs away. Therefore, we combine both losses for better optimisation (see below for ablation results).

Figure \ref{fig:loss} explains how our approach proposes this dual loss optimisation. We address query-title pairs where semantic distance is not proportional to the centrality specifications defined by previously annotated data. As can be seen from the figure, dual loss optimisation ensures that for each query ($Q$), the maximum intra-class distance (blue arrows) is smaller than the minimum inter-class distance (the red arrow). We define a radius/margin \textbf{$m$}, for all the central product titles, while all the non-central product titles fall outside the margin. Please note that the loss penalises the model for non-central titles \mbox{having a distance to $Q$ less than \textbf{$m$}.}





\begin{table}[h!]
    \centering
    \resizebox{0.9\columnwidth}{!}{%
    \begin{tabular}{c|c|c|c}
    
    \toprule
    \textbf{Eval Split} & \textbf{\# Corpus} & \textbf{\# Dev-Q} & \textbf{\# Test-Q} \\
    \midrule
    {\it CQ} & $187469$ & $5776$ & $17325$ \\
    {\it CQ-balanced} & $46561$ & $5776$ & $17325$ \\
    {\it CQ-common-str} & $12508$ & $2117$ & $6351$ \\
    {\it CQ-alphanum} & $162115$ & $4111$ & $12333$ \\
    \bottomrule
    \end{tabular}%
    }
    \caption{Data Distribution in each split. Q -> queries}
    \vspace{-0.5cm}
    \label{tab:eval_splits}
\end{table}

\section{Experiment Setup}
\begin{table*}[t!]
    \centering
    \resizebox{0.95\textwidth}{!}{%
    \begin{tabular}{c|c|ccc|ccc|ccc|c}
    \toprule
    \textbf{Encoder} & \textbf{UCO} & \multicolumn{3}{c}{$\textbf{Precision}@k~(\uparrow)$}&\multicolumn{3}{c}{$\textbf{Recall}@k~(\uparrow)$}&\multicolumn{3}{c}{$\textbf{NDCG}@k~(\uparrow)$}&\textbf{MRR} ($\uparrow$)\\ 
    & & $\rm{\textbf{3}}$&$\rm{\textbf{5}}$&$\rm{\textbf{10}}$& $\rm{\textbf{3}}$&$\rm{\textbf{5}}$&$\rm{\textbf{10}}$& $\rm{\textbf{3}}$&$\rm{\textbf{5}}$&$\rm{\textbf{10}}$&$\rm{\textbf{@10}}$\\
    
    \midrule\midrule
    \multicolumn{12}{c}{\textbf{\textit{CQ} test}} \\
    \midrule
    \multirow{1}{*}{BERT}& \xmark & $16.20$ & $13.03$ & $8.93$ & $11.31$
    & $14.41$ & $18.83$ & $0.1912$ & $0.1818$ & $0.1833$ & $0.2771$ \\
    \midrule
    \multirow{2}{*}{eBERT}& \xmark & $20.71$ & $17.25$ & $12.54$ & $14.46$ & $19.19$ & $26.26$ & $0.2392$ & $0.2330$ & $0.2430$ & $0.3415$\\
     & \cmark & \cellcolor{blue!20} $64.76$ & \cellcolor{blue!20}$55.74$ & \cellcolor{blue!20}$39.22$ & \cellcolor{blue!20}$49.63$ & \cellcolor{blue!20}$63.92$ & \cellcolor{blue!20}$79.65$ & \cellcolor{blue!20}$0.7439$ & \cellcolor{blue!20}$0.7488$ & \cellcolor{blue!20}$0.7672$ & \cellcolor{blue!20}$0.8189$ \\
    \midrule
    \multirow{2}{*}{\shortstack{eBERT \\ (siam)}}& \xmark & $55.25$ & $48.33$ & $34.90$ & $42.36$ & $56.09$ & $72.22$ & $0.6315$ & $0.6428$ & $0.6704$ & $0.7263$\\
    & \cmark &\cellcolor{blue!20} $66.25$ & \cellcolor{blue!20}$57.16$ & \cellcolor{blue!20}$40.20$ & \cellcolor{blue!20}$51.18$ & \cellcolor{blue!20}$65.79$ & \cellcolor{blue!20}$81.66$ & \cellcolor{blue!20}$0.7635$ & \cellcolor{blue!20}$0.7698$ & \cellcolor{blue!20}$0.7886$ & \cellcolor{blue!20}$0.8347$\\
    \midrule
    \multicolumn{12}{c}{\textbf{\textit{CQ-balanced} test}} \\
    \midrule
    \multirow{1}{*}{BERT}& \xmark & $7.13$ & $4.94$ & $2.95$ & $21.26$ & $24.58$ & $29.33$ & $0.1824$ & $0.1961$ & $0.2115$ & $0.1862$\\
    \midrule
    \multirow{2}{*}{eBERT}& \xmark & $9.72$ & $6.94$ & $4.22$ & $29.02$ & $34.58$ & $42.07$ & $0.2428$ & $0.2657$ & $0.2899$ & $0.2495$\\
     & \cmark & \cellcolor{blue!20} $28.57$ & \cellcolor{blue!20}$18.15$ & \cellcolor{blue!20}$9.50$ & \cellcolor{blue!20}$85.40$ & \cellcolor{blue!20}$90.42$ & \cellcolor{blue!20}$94.62$ & \cellcolor{blue!20}$0.7851$ & \cellcolor{blue!20}$0.8059$ & \cellcolor{blue!20}$0.8197$ & \cellcolor{blue!20}$0.7789$ \\
    \midrule
    \multirow{2}{*}{\shortstack{eBERT \\ (siam)}}& \xmark & $25.99$ & $16.68$ & $8.89$ & $77.66$ & $83.08$ & $88.59$ & $0.6888$ & $0.7112$ & $0.7291$ & $0.6784$\\
    & \cmark & \cellcolor{blue!20}$29.19$ & \cellcolor{blue!20}$18.39$ & \cellcolor{blue!20}$9.58$ & \cellcolor{blue!20}$87.26$ & \cellcolor{blue!20}$91.58$ & \cellcolor{blue!20}$95.43$ & \cellcolor{blue!20}$0.8046$ & \cellcolor{blue!20}$0.8225$ & \cellcolor{blue!20}$0.8351$ & \cellcolor{blue!20}$0.7965$\\
    \midrule
    \multicolumn{12}{c}{\textbf{\textit{CQ-common-str} test}} \\
    \midrule
    \multirow{1}{*}{BERT}& \xmark & $9.41$ & $6.31$ & $3.65$ & $28.15$ & $31.47$ & $36.35$ & $0.2532$ & $0.2669$ & $0.2828$ & $0.2579$ \\
    \midrule
    \multirow{2}{*}{eBERT}& \xmark & $12.62$ & $8.64$ & $5.00$ & $37.79$ & $43.10$ & $49.92$ & $0.3272$ & $0.3491$ & $0.3714$ & $0.3315$ \\
     & \cmark & \cellcolor{blue!20} $32.03$ & \cellcolor{blue!20}$19.58$ & \cellcolor{blue!20}$9.92$ & \cellcolor{blue!20}$95.84$ & \cellcolor{blue!20}$97.65$ & \cellcolor{blue!20}$98.87$ & \cellcolor{blue!20}$0.9091$ & \cellcolor{blue!20}$0.9166$ & \cellcolor{blue!20}$0.9206$ & \cellcolor{blue!20}$0.8979$ \\
    \midrule
    \multirow{2}{*}{\shortstack{eBERT \\ (siam)}}& \xmark & $29.93$ & $18.76$ & $9.68$ & $89.57$ & $93.58$ & $96.50$ & $0.8194$ & $0.8361$ & $0.8456$ & $0.8063$\\
    & \cmark & \cellcolor{blue!20} $32.12$ & \cellcolor{blue!20}$19.64$ & \cellcolor{blue!20}$9.92$ & \cellcolor{blue!20}$96.11$ & \cellcolor{blue!20}$97.94$ & \cellcolor{blue!20}$98.93$ & \cellcolor{blue!20}$0.9117$ & \cellcolor{blue!20}$0.9193$ & \cellcolor{blue!20}$0.9226$ & \cellcolor{blue!20}$0.9003$\\
    \midrule
    \multicolumn{12}{c}{\textbf{\textit{CQ-alphanum} test}} \\
    \midrule
    \multirow{1}{*}{BERT}& \xmark & $20.54$ & $16.65$ & $11.47$ & $13.45$ & $17.32$ & $22.82$ & $0.2333$ & $0.2176$ & $0.2226$ & $0.3350$ \\
    \midrule
    \multirow{2}{*}{eBERT}& \xmark & $23.35$ & $19.54$ & $13.77$ & $15.53$ & $20.76$ & $27.85$ & $0.2630$ & $0.2516$ & $0.2617$ & $0.3739$ \\
    & \cmark & \cellcolor{blue!20}$64.58$ & \cellcolor{blue!20}$57.27$ & \cellcolor{blue!20}$40.35$ & \cellcolor{blue!20}$44.05$ & \cellcolor{blue!20}$59.97$ & \cellcolor{blue!20}$77.00$ & \cellcolor{blue!20}$0.7119$ & \cellcolor{blue!20}$0.7094$ & \cellcolor{blue!20}$0.7344$ & \cellcolor{blue!20}$0.8018$ \\
    \midrule
    \multirow{2}{*}{\shortstack{eBERT \\ (siam)}}& \xmark & $60.67$ & $54.10$ & $38.54$ & $41.32$ & $57.10$ & $74.20$ & $0.6652$ & $0.6654$ & $0.6951$ & $0.7618$\\
    & \cmark & \cellcolor{blue!20} $67.10$ & \cellcolor{blue!20}$59.70$ & \cellcolor{blue!20}$41.81$ & \cellcolor{blue!20}$46.07$ &\cellcolor{blue!20} $62.72$ & \cellcolor{blue!20}$79.76$ & \cellcolor{blue!20}$0.7375$ & \cellcolor{blue!20}$0.7371$ & \cellcolor{blue!20}$0.7609$ & \cellcolor{blue!20}$0.8171$\\
    \bottomrule
    \end{tabular}%
    }
    \caption{Evaluating the efficacy of the proposed UCO on the all test sets, using different encoder backbones. Precision and Recall values are shown in ($\%$); higher values are preferred.}
    \label{tab:mainresults}
    \vspace{-0.3cm}
\end{table*}

\subsection{Dataset Curation}
We preprocess all query-title pairs from the IGR dataset by filtering out non-English pairs to ensure linguistic consistency and relevance.
Once preprocessed, we select queries that have both the corresponding positive titles (relevancy > $3$) and negative titles (relevancy < $3$) from the IGR dataset. This selection forms our initial split, referred to as {\it Common Queries (CQ)}. We observed a notable imbalance towards positive query-title pairs in {\it CQ}, stemming from the inherent nature of e-commerce product listings and the data collection strategy highlighted in Section 1, which emphasises capturing relevant matches. To address this imbalance and ensure a fair comparison, we introduce a balanced version of {\it CQ}, where the number of positive and negative query-product title pairs is approximately equal, referred to as {\it CQ-balanced}.

\begin{figure}[h]
    \centering
    \begin{subfigure}[b]{\columnwidth}
        \centering
        \includegraphics[width=\textwidth]{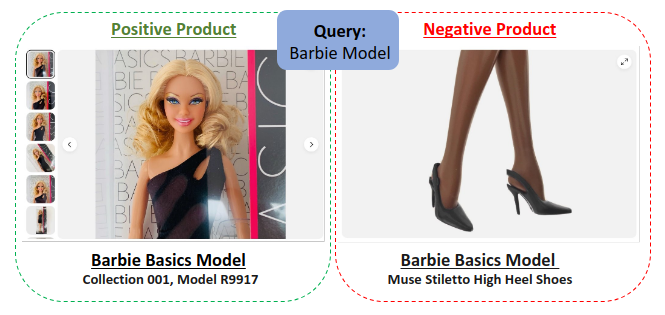}
        \caption{The sub-string ``Barbie Model'' is a part of both positive and negative product titles.}
        \label{fig:samplesub1}
    \end{subfigure}
    \hfill
    \begin{subfigure}[b]{\columnwidth}
        \centering
        \includegraphics[width=\textwidth]{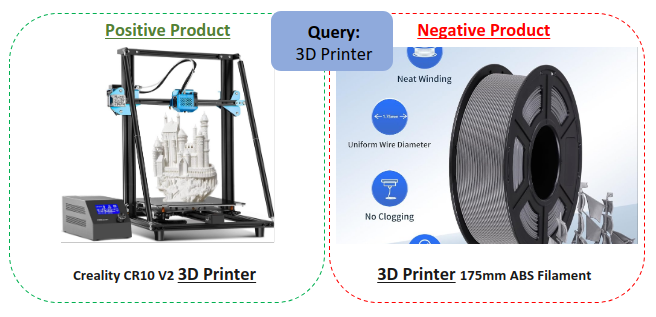}
        \caption{The sub-string ``3D Printer'' is a part of both positive and negative product titles.}
        \label{fig:samplesub2}
    \end{subfigure}
    \caption{Examples of query-title pairs from the {\it CQ-common-str} split. Both, positive and negative product titles have high semantic correlation to the user query, however only the positive product title exhibits a central idea/intent.}
    \label{fig:cq_str_ex}
\end{figure}

Upon examining the query-title pairs, as also discussed in Section 1, we found that often, the exact string of a query appears in both positive and negative product titles. We isolate these query-title pairs to form our third split, named {\it CQ-common-str} (see Figure \ref{fig:cq_str_ex}). This task necessitates considering both, user centrality and semantic connections between the query and product titles. We conduct a correlation test, and observe that Pearson, Kendall and Spearman correlations between the \textit{graded relevance score} and the \textit{binary centrality score} are $0.78$, $0.73$ and $0.77$, respectively,  validating our assumption that both types of scores are highly correlated and hence the ranked results are expected to conform with the overall pattern of the dataset.

Lastly, to facilitate the evaluation of our proposed methodology specifically on alphanumeric query-title pairs, we create a separate split containing only queries and titles with alphanumeric characters, referred as {\it CQ-alphanum}. For each evaluation split, all the positive and negative titles constitute the retrieval corpus, while we create distinct development and test query sets in an 80:20 ratio. Table \ref{tab:eval_splits} shows the number of entries in the corpus and query sets for each split. The development query set assists in selecting the best-performing UCO model (\textit{i.e.,} during optimisation on user-intent centrality), while the unseen test query set validates the ranking capability of UCO.




\paragraph{Visual Samples}
\label{secapp:visamp}

We believe that the split, {\it CQ-common-str}, presents the most demanding evaluation scenario, requiring the model to simultaneously differentiate the semantic relationships of the strings in both positive and negative product titles. 

Figure~\ref{fig:samplesub1} shows common query string in positive and negative titles:  query ``barbie model" , positive title ``Barbie Top Model Summer Doll 2008 Ginger Hair" (the real doll) , negative title ``Barbie Model Pointed Toe Fashion High Heel Shoes" (only the shoes). Similarly, Figure~\ref{fig:samplesub2} shows the query ``3d printer", positive title ``Creality CR10 V2 3D Printer", and the negative title ``3D Printer 175mm ABS Filament Made in UAE Premium Quality"; where the negative title is just the printer filaments. Note that query-title pairs such as these are challenging for traditional IR methods too, which use lexical matching.

\subsection{Implementation Details}

We optimise both the encoder backbones on the centrality score classification-train split for a maximum of $10$ epochs. During  training, we  run two sequential evaluators on both the centrality scores and the retrieval ranking  in the curated IGR datasets. First, an evaluator that will compute the embeddings for both query and title and use them to calculate the cosine similarity. If the similarity is above a threshold, we have a central title. Second, given a query and the corpus of all titles, the evaluator finds the most relevant product title to the query (top $3$, $5$ and $10$ titles). During optimisation, we save the checkpoint that performs best on the second evaluator. For all experiments, we use a batch size of $32$, with the Adam optimiser and $2e-05$ as the learning rate, and $0.01$ as weight decay. Optimising one encoder backbone using the above parameters takes $~30$ hours on a single NVIDIA V100 GPU. For evaluation, we use cosine similarity as scoring function.

\paragraph{Evaluation Metrics} We use different existing evaluation metrics to measure the overall model performance. Precision@$k$ measures the proportion of relevant products in the top-$k$ recommendations (considering their relevance), while Recall@$k$ measures the proportion of relevant products that were retrieved among all relevant products (irrespective of their rank). NDCG~\cite{10.1145/582415.582418} measures the ranking quality by comparing the recommended items' order against an ideal ranking. As a result, NDCG considers both the relevance and rank of the recommended products. Mean Reciprocal Rank (MRR) evaluates the average rank of the first relevant item across all queries. A high MRR is an indication of being able to provide users with relevant products ranked as high as possible.



\section{Results and Discussion}
Considering various aspects like retrieval and ranking quality, we analyse model performance using a diverse set of metrics (explained in $\S$\ref{secapp:visamp}). We also perform an ablation test on the eBERT model to identify the contribution of both loss functions, MNRL and OCL, and discuss the qualitative analysis below. 
Table~\ref{tab:mainresults} displays the results for each of the evaluation splits, {\it CQ}, {\it CQ-balanced}, {\it CQ-common-str} and {\it CQ-alphanum}. Across each split, a consistent pattern emerges: the incorporation of UCO leads to a substantial improvement in product retrieval performance across all metrics. This improvement is evident regardless of whether the backbone encoder employed is eBERT or eBERT-siam. This highlights UCO's capability to enhance an existing model's embedding space, enabling it to capture semantic relationships between user queries and product titles attuned to the user intent, thus retrieving products with high user centrality. It is evident that BERT, a publicly available model, was unable to capture query-title relations given it was not pre-trained on internal data. Even with internal models, the results without UCO show the challenge posed by these evaluation splits curated for this work. For alphanumeric queries, the NDCG performance improvement ranges from $7\%$ points for the base model to $47\%$ points, including the model fine-tuned with the Siamese approach, demonstrating the efficacy of UCO. For query-titles with common strings, it ranges from $8\%$ to $58\%$ points. We also see similar improvements in all metrics, for the other two evaluation sets. 

\paragraph{Loss Ablation} We conducted a quick ablation test over the \textit{CQ} evaluation split. For this test, we fine-tuned the eBERT and eBERT-siam models using individual loss functions and their combination, which is our finalised approach. From Table~\ref{tab:ablation_loss}, it is clear that the combination of both loss functions helps improve performance for both models. We evaluate this using both NDCG and MRR evaluation metrics. When employed individually, MNRL seems to outperform OCL in both metrics. Overall, dual-loss based optimisation emerges as a clear winning strategy. 

\paragraph{Qualitative Analysis} We discuss the performance improvement shown by UCO with two examples in Figures~\ref{fig:qual-cqstr} and ~\ref{fig:qual-alphanum}, \textit{shown in the Appendix below}. We use the eBERT-siamese model to rank retrieved products with and without UCO optimisation. In Figure~\ref{fig:qual-cqstr}, search query `1080' from the test set retrieves more `central' products when UCO optimised model is used, \textit{i.e.,} graphics card variants. Similarly, on the use of the alphanumeric search query in Figure~\ref{fig:qual-alphanum}, most relevant products are ranked on top, \textit{i.e.,} keyboard with the same product identifier, showing how UCO model optimisation helps rank relevant products on top. 

\begin{table}[t]
    \centering
    \resizebox{\columnwidth}{!}{%
    \begin{tabular}{c|cc|cc}
    \toprule
    \multirow{2}{*}{\textbf{Loss}} & \multicolumn{2}{c}{\textbf{eBERT}} & \multicolumn{2}{c}{\textbf{eBERT-siam}}\\ 
    & \rm{$\textbf{NDCG@5}$} & \rm{$\textbf{MRR@10}$} & \rm{$\textbf{NDCG@5}$} & \rm{$\textbf{MRR@10}$} \\
    \midrule
    MNRL &  $0.7139$ & $0.7899$ & $0.7254$ & $0.8016$\\
    OCL & $0.5497$ & $0.6559$ & $0.5812$ & $0.6978$\\
    \midrule
    \textbf{MNRL + OCL} & \cellcolor{blue!20} $0.7488$ & \cellcolor{blue!20}$0.8189$ & \cellcolor{blue!20}$0.7698$ & \cellcolor{blue!20}$0.8347$ \\
    
    \bottomrule
    \end{tabular}%
    }
    \caption{Ablation experiment to study the efficacy of MNRL and OCL losses when taken individually; higher values are preferred.}
    \vspace{-0.5cm}
    \label{tab:ablation_loss}
\end{table}




\section{Conclusion and Future Work}

This work addresses product search queries that represent an important challenge for e-commerce platforms. The main challenge occurs when the retrieved titles are semantically relevant, but not \textit{central} to the user-intent as is reflected by the specificity of the query. The challenge is even greater with ambiguous queries where the same query string is present in both relevant and irrelevant titles as well as when queries are alphanumeric. We address the semantic complexity of these challenging query-title pairs by fine-tuning existing internal models with a user-intent centrality optimisation (UCO) step to infuse information about the typicality of query-title pairs. The retrieval model performance showed significant improvement with several hard example datasets with a dual-loss based optimisation approach, which pays attention to negative pairs that have a lower distance than the positive pairs with the largest distance. The dual-loss based optimisation helps in separating the irrelevant pairs of queries and titles while keeping the distance smaller for relevant query-title pairs. The improvement in ranking performance demonstrated by our approach helps identify and categorise what users intend to find online when they search the platform.

In future, we aim to restructure queries in our hard-negative pairs to be less ambiguous. Leveraging GenAI-based prompt engineering and explainability using approaches like chain-of-thought, we can investigate titles that indicate typical queries, \textit{aligning them closer to the user intent}, and moving towards \textit{explainable product retrieval}.



\bibliography{custom}

\clearpage

\appendix

\begin{figure*}[ht]
    \centering
    \includegraphics[width=0.8\textwidth]{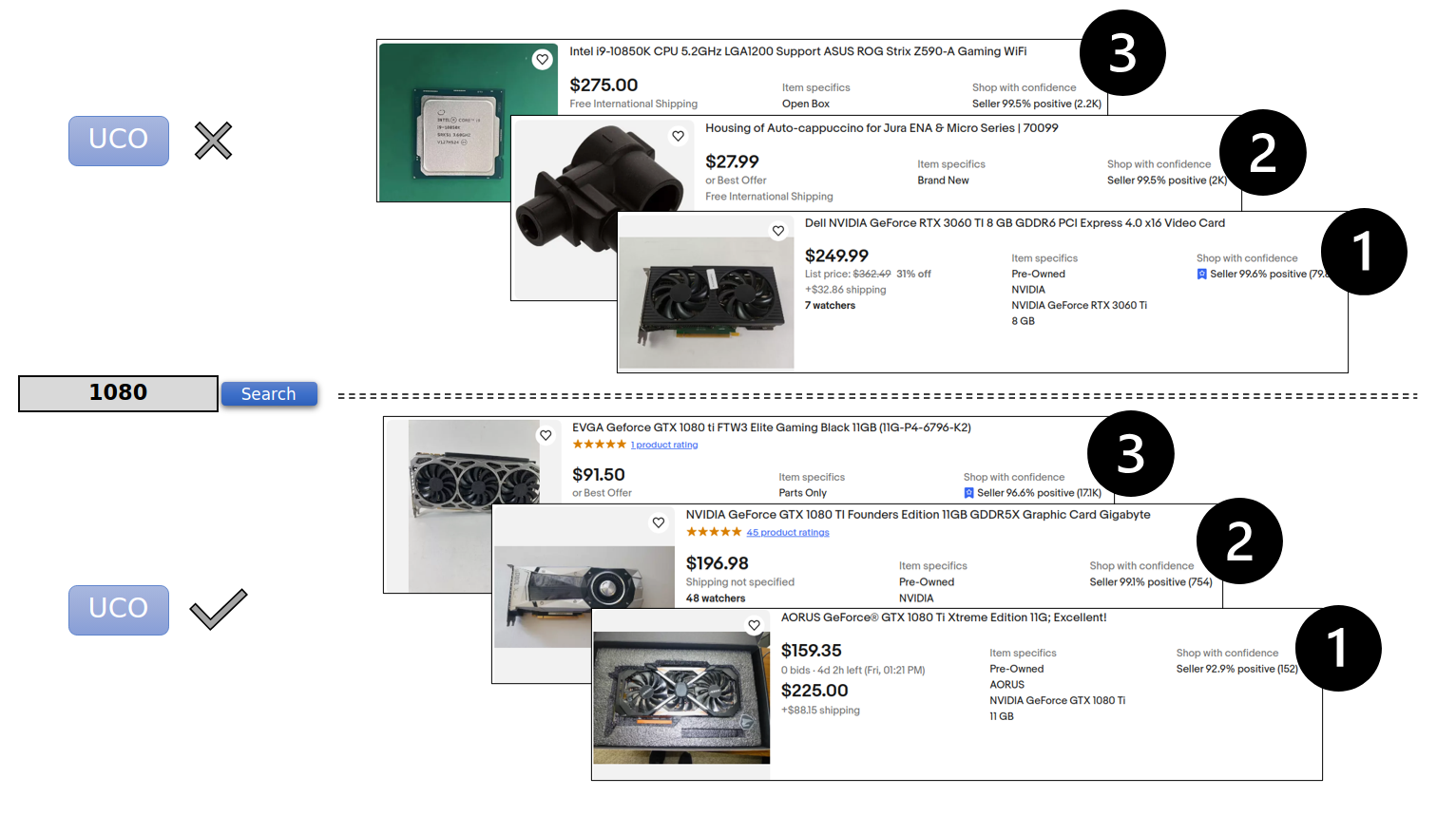}
    \caption{Qualitative comparison of the proposed UCO on a sample from the \textit{CQ-common-str} test set, when using the eBERT (siam) as the encoder backbone. We showcase the top-3 retrieved product titles for both encoders.}
    \label{fig:qual-cqstr}
\end{figure*}

\begin{figure*}[ht]
    \centering
    \includegraphics[width=0.8\textwidth]{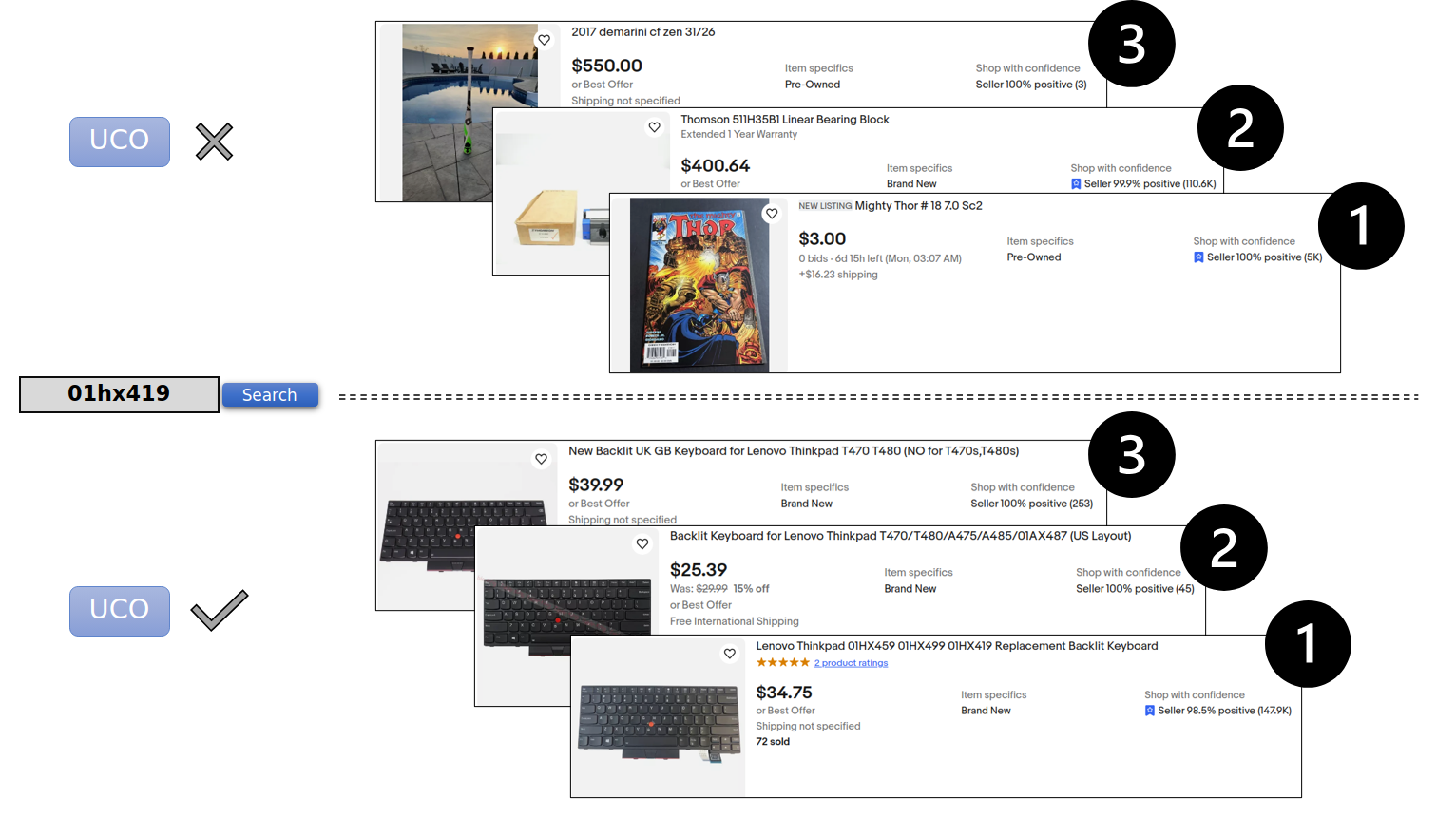}
    \caption{Qualitative comparison of the proposed UCO on a sample from the \textit{CQ-alphanum} test set, when using the eBERT (siam) as the encoder backbone. We showcase the top-3 retrieved product titles for both encoders.}
    \label{fig:qual-alphanum}
\end{figure*}

\end{document}